\documentclass[graybox]{svmult}

\usepackage{mathptmx}       
\usepackage{helvet}         
\usepackage{courier}        
\usepackage{type1cm}        
\usepackage{makeidx}         
\usepackage{graphicx}        
\usepackage{multicol}        
\usepackage[bottom]{footmisc}
\usepackage{subfigure}
\usepackage{amsmath}

\makeindex             

\begin{document}

\title*{Die-out Probability in SIS Epidemic Processes on Networks}
\author{Qiang Liu and Piet Van Mieghem}
\institute{Faculty of Electrical Engineering, Mathematics and Computer Science \at Delft University of Technology, P.O Box 5031, 2600 GA Delft, The Netherlands\\ \email{\{Q.L.Liu, P.F.A.VanMieghem\}@TuDelft.nl}}
%
%
\maketitle

\abstract*{An accurate approximate formula of the die-out probability in a SIS epidemic process on a network is proposed. The formula contains only three essential parameters: the largest eigenvalue of the adjacency matrix of the network, the effective infection rate of the virus, and the initial number of infected nodes in the network. The die-out probability formula is compared with the exact die-out probability in complete graphs, Erd\H{o}s-R\'enyi graphs, and a power-law graph. Furthermore, as an example, the formula is applied to the $N$-Intertwined Mean-Field Approximation, to explicitly incorporate the die-out.}

\abstract{An accurate approximate formula of the die-out probability in a SIS epidemic process on a network is proposed. The formula contains only three essential parameters: the largest eigenvalue of the adjacency matrix of the network, the effective infection rate of the virus, and the initial number of infected nodes in the network. The die-out probability formula is compared with the exact die-out probability in complete graphs, Erd\H{o}s-R\'enyi graphs, and a power-law graph. Furthermore, as an example, the formula is applied to the $N$-Intertwined Mean-Field Approximation, to explicitly incorporate the die-out.}

\section{Introduction}
The SIS epidemic process models spreading phenomena of information or viruses on networks \cite{pastor-satorras_epidemic_2015}. In a network, each node has two states: susceptible and infected. A Bernoulli random variable $X_j(t)\in \{0,1\}$ denotes the state of each node, where $X_j(t)=0$ means that node $j$ is susceptible and $X_j(t)=1$ indicates that node $j$ is infected at time $t$. An infected node can infect its susceptible neighbors with a infection rate $\beta$ by changing the susceptible neighbor nodes into infected nodes, and each infected node is cured and becomes a susceptible node with a curing rate $\delta$. If the infection and curing processes are Poisson processes, the SIS epidemic model is Markovian, where the sojourn times in the infected and susceptible state are exponentially distributed. The governing equation of a node $j$ in the Markovian SIS epidemic process in an unweighted and undirected network with N nodes, represented by an $N\times N$ symmetric adjacency matrix $A$, is \cite[p.~449]{van_mieghem_performance_2014}
\begin{equation}\label{eq_SIS_governing_equation}
\frac{\D E[X_j(t)]}{dt}=-\delta E[X_j(t)]+\beta\sum_{k=1}^{N}a_{kj}E[X_k(t)]-\beta\sum_{k=1}^{N}a_{kj}E[X_j(t)X_k(t)]
\end{equation}

The epidemic threshold $\tau_c$ of the SIS epidemic process implies that, if the effective infection rate $\tau=\beta/\delta>\tau_c$, the virus will spread over the network for a very long time, and if $\tau<\tau_c$, the number of infected nodes decreases exponentially fast after sufficiently long time \cite{pastor-satorras_epidemic_2015,van_mieghem_approximate_2016}. There is an approximate value \cite{wang_epidemic_2003-1} and lower bound \cite{van_mieghem_virus_2009} of the epidemic threshold $\tau_c>\tau_c^{(1)}=1/\lambda_1$, where $\lambda_1$ is the largest eigenvalue of the adjacency matrix $A$. In this paper, the threshold $\tau_c^{(1)}$ is referred to as the $N$-Intertwined Mean Field Approximation (NIMFA) threshold, where the superscript $(1)$ in $\tau_c^{(1)}$ refers to the fact that NIMFA is a first order mean-field approximation \cite{van_mieghem_virus_2009}.



The structure of this paper is organized as follows. Section \ref{sec_The_prevalence_and_the_die_out_probability} introduces the relation between the prevalence (\ref{eq_prevalence}) and the average fraction of infected nodes conditioned to the survival of the virus. Clearly, the virus die-out probability plays a key role. Section \ref{sec_the_die_out_probability_of_virus_an_accurate_approximation} proposes an accurate approximate formula (\ref{eq_formula_of_die_out_probability}) for the die-out probability in the metastable state of the SIS epidemic process. Figure \ref{fig_die-out_probability} and Fig. \ref{fig_randomgraphdieout} demonstrate the accuracy and the limitation of (\ref{eq_formula_of_die_out_probability}) in complete graphs, Erd\H{o}s-R\'enyi graphs, and power-law graphs. Finally, we apply formula (\ref{eq_formula_of_die_out_probability}) to correct the NIMFA prevalence (\ref{eq_NIMFA_prevalence}) as shown in Fig. \ref{fig_NIMFA_correct}.

\section{The Prevalence and the Die-out Probability}
\label{sec_The_prevalence_and_the_die_out_probability}
The prevalence $y(t)$ of a SIS epidemic process is the expected fraction of infected nodes at time $t$,
\begin{equation}\label{eq_prevalence}
  y(t)=E[S(t)]
\end{equation}
where $S(t)=\frac{1}{N}\sum_{j=1}^{N}X_j(t)$ is the fraction of infected nodes. The prevalence in the exact Markovian epidemic process after infinitely long time tends to zero, where the absorbing state is reached. Before the virus dies out, virus may exist in networks for a very long time \cite{van_de_bovenkamp_survival_2015,van_mieghem_decay_2013}. In the metastable state, the prevalence $y(t)$ changes very slowly and there is a balance between the infection and curing processes. We confine ourselves to the time region $[0, t_{max}]$ that the prevalence $y(t_{max})\neq 0$, and the prevalence is approximately equal at every time $t\in [t_m,t_{max}]$, where $t_m$ is the time that the SIS process reaches metastable state.
However, for one realization of the epidemic process, we cannot expect that the fraction of infected nodes oscillates around the level of the prevalence $y(t)$ with time $t$, because the prevalence $y(t)$ is the average over all possible realizations including the die-out realizations. In real observed diseases, the virus has not died out yet, so that the fraction of infected population is positive. So, there are actually two kinds of average: the average over all possible realizations (prevalence), and the average over the realizations conditioned to the survival of the virus. To prevent confusion, the faction of infected nodes under the condition that the virus survives at time $t$ is denoted by a random variable $\tilde{S}(t)$ in this paper. Consequently, we have $\Pr[\tilde{S}(t)=i/N]=\Pr[S(t)=i/N|S(t)\neq 0]$ for $\tilde{S}(t)\in\{1/N,2/N,\cdots,1\}$ while $S(t)\in\{0,1/N,2/N,\cdots,1\}$. The removal of the absorbing state \cite{cator_susceptible-infected-susceptible_2013} or the assumption that the virus survives is associated with the quasi-stationarity or metastability of the SIS process \cite{sander_sampling_2016}. The expectation of $\tilde{S}(t)$ of an epidemic process in a network with $N$ nodes is
$$
  E[\tilde{S}(t)] = \sum_{i=1}^{N}\frac{i}{N}\Pr[\tilde{S}(t)]= \sum_{i=1}^{N}\frac{i}{N}\Pr\left[\left.S(t)=\frac{i}{N}\right|S(t)\neq 0\right]
$$
With the definition of the conditinal probability,
\begin{eqnarray*}
	\Pr\left[\left.S(t)=\frac{i}{N}\right|S(t)\neq 0\right]&=& \frac{\Pr\left[\left\{S(t)=\frac{i}{N}\right\}\cap\left\{S(t)\neq 0\right\}\right]}{\Pr[S(t)\neq 0]} \\\nonumber
            				&=& \frac{\Pr\left[S(t)=\frac{i}{N}\right]}{\Pr\left[S(t)\neq 0\right]}\ \  \mbox{provided} \ i>0\nonumber
\end{eqnarray*}
we have
$$
    E[\tilde{S}(t)]= \frac{1}{\Pr[S(t)\neq 0]}\sum_{i=0}^{N}\frac{i}{N}\Pr\left[S(t)=\frac{i}{N}\right]
                   = \frac{E[S(t)]}{\Pr[S(t)\neq 0]}\nonumber
$$
Since $\Pr[S(t)\neq 0]=1-\Pr[S(t)=0]$, the prevalence can be written as
\begin{equation}\label{eq_S_and_Sq}
	y(t)=\tilde{y}(t)\left(1-\Pr[S(t)= 0]\right)
\end{equation}
where $\tilde{y}(t)=E[\tilde{S}(t)]$. Equation (\ref{eq_S_and_Sq}) shows the relation between the prevalence $y(t)$ and the average fraction $\tilde{y}(t)$ of infected nodes under the condition that the virus survives, where the die-out probability $\Pr[S(t)=0]$ is essential. Both the prevalence $y(t)$ and the virus die-out probability $\Pr[S(t)=0]$ are difficult to compute analytically in general graphs.

The Markovian SIS epidemic process on the complete graph $K_N$ is a birth-and-death process \cite{cator_susceptible-infected-susceptible_2013,van_mieghem_performance_2014}. The states $\{0,1,\cdots, N\}$ of the birth-and-death process are the number of infected nodes, where $0$ is the absorbing state or overall-healthy state. Therefore, the die-out probability $\Pr[S(t)=0]$ can be obtained by solving the birth-and-death process,
\begin{equation}\label{eq_birth_death_markov_chain_differential_equation}
\left(\vec{s}'(t)\right)^T=\vec{s}^T(t)\tens{Q}
\end{equation}
where $\tens{Q}$ is the infinitesimal generator of the birth-and-death Markov chain, and $\vec{s}^T(t)=[s_0(t),\cdots,s_N(t)]$ is the state probability vector with each element $s_i(t)=\Pr[S(t)=i/N]$ for $0\leq i \leq N$, and $s_0(t)=\Pr[S(t)=0]$.

The die-out probability $\Pr[S(t)=0]$ of SIS epidemic process in complete graphs also equals the gambler's ruin probability \cite[p.~231]{van_mieghem_performance_2014} as shown in the Appendix,
\begin{equation}\label{eq_gambler_ruin_with_polynomial}
\mu_{n}=\frac{\sum_{j=0}^{N-n-1}j!\tau^{j}}{\sum_{j=0}^{N-1}j!\tau^{j}}
\end{equation}
Different from solving (\ref{eq_birth_death_markov_chain_differential_equation}), Eq. (\ref{eq_gambler_ruin_with_polynomial}) only applies to the metastable state and cannot be used to calculate the die-out probability at an arbitrary time $t$. As demonstrated in the Appendix, Eq.(\ref{eq_gambler_ruin_with_polynomial}) upper bounds the actual die-out probability, because (\ref{eq_gambler_ruin_with_polynomial}) assumes that the virus wins only when it infects all $N$ nodes in a finite time before dying out.

\section{The Die-out Probability: an Accurate Approximation}
\label{sec_the_die_out_probability_of_virus_an_accurate_approximation}
Apart from solving (\ref{eq_birth_death_markov_chain_differential_equation}) or employing the gambler's ruin formula (\ref{eq_gambler_ruin_with_polynomial}), in this section we propose a novel approximate formula of the virus die-out probability in the metastable state.

We assume that the prevalence $y(t)$ is approximately constant in the metastable state. Relation (\ref{eq_S_and_Sq}) then indicates that the die-out probability is also approximately constant. In the metastable state, we then find that the virus die-out probability in a sufficiently large graph is approximately
\begin{equation}\label{eq_formula_of_die_out_probability}
  \Pr[S(t_m)=0]\approx\frac{1}{x^n},\ \ \mbox{with}\ x\geq 1
\end{equation}
where $S(t_m)$ denotes the fraction of infected nodes of the SIS epidemic process in the metastable reached at time $t_m$, $x=\tau/\tau_c^{(1)}=\lambda_1 \tau$ is the normalized effective infection rate of the virus, and $n$ is the number of initially infected nodes. The situation $x<1$ is not considered, because the infection rate is below the threshold and the SIS process dies out before reaching the metastable state. In addition, $1/x>1$ cannot represent a probability. As the first order NIMFA threshold $\tau_c^{(1)}=1/\lambda_1$ is a lower bound of the actual threshold $\tau_c$, the prevalence $y(t)$ decreases exponentially fast for sufficiently large time \cite{van_mieghem_approximate_2016} when $x\leq 1$, and the virus die-out probability tends to $1$. Also, the accuracy of formula (\ref{eq_formula_of_die_out_probability}) is related to the accuracy of the NIMFA threshold $\tau_c^{(1)}=1/\lambda_1$. For example, if the effective infection rate is below the real threshold and $\tau_c^{(1)}<\tau<\tau_c$, formula $1/x^n<1$, but the virus dies out within finite time with the probability tending to $1$. In the Appendix, an analytically approach to (\ref{eq_formula_of_die_out_probability}) from the gambler's ruin probability (\ref{eq_gambler_ruin_with_polynomial}) in complete graphs is presented.

By introducing the normalized effective infection rate $x=\tau/\tau_c^{(1)}=\tau\lambda_1$ into (\ref{eq_formula_of_die_out_probability}), the network topology---the largest eigenvalue of the adjacency matrix $\lambda_1$---is reflected. Formula (\ref{eq_formula_of_die_out_probability}) is simple, and only three essential parameters are involved: the spectral radius $\lambda_1$, the virus spreading ability $\tau$, and the initially infected number of nodes $n$. If a few nodes are infected and the infection rate is above the threshold $x>1$, then formula (\ref{eq_formula_of_die_out_probability}), which is equivalent to $\Pr[S(t_m)=0]\approx \mathrm{e}^{-n\log x}$, shows that the network will experience a disease outbreak, because the die-out probability decreases exponentially fast with $n$ above the epidemic threshold ($\log x>0$).

In the sequel, we compare $(\ref{eq_formula_of_die_out_probability})$ and the die-out probability $\Pr[S(t)=0]$ obtained via simulations. The curing rate of all the calculations and simulations below is $\delta=1$.
\subsection{Complete Graphs}
\begin{figure}[t]
    \subfigure[The virus die-out probability and the prevalence of epidemic process in complete graph $K_{126}$. Initially $3$ nodes are infected. This figure shows a clearly metastable state region.]{
      \centering
      \includegraphics[width=152pt]{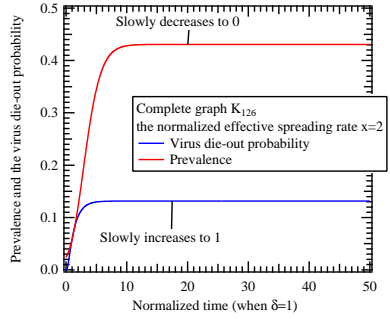}
      \label{K126example}
    }
    \qquad
    \centering
    \subfigure[The die-out probabilities from simulation of the SIS epidemic process and calculation of the birth-and-death process are shown with $1$ initially infected node. With the increase of network size $N$, the die-out probabilities converge to the simple formula: $1/x^n$.]{
      \centering
      \includegraphics[width=152pt]{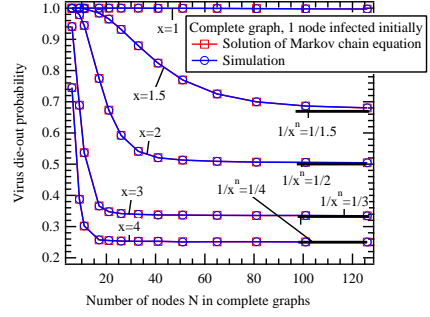}
      \label{one_initial_infected}
    }
    \\
    \subfigure[With $2$ nodes infected initially, this figure verifies (\ref{eq_formula_of_die_out_probability}) as Fig. \ref{one_initial_infected} with simulation and calculation results.]{
      \centering
      \includegraphics[width=152pt]{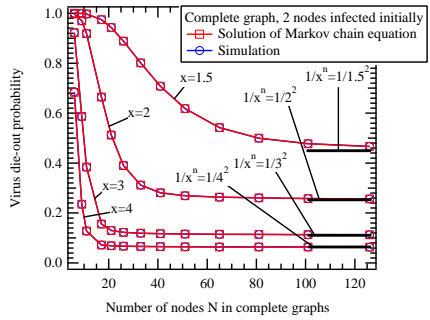}
      \label{two_initial_infected}
    }
    \qquad
    \subfigure[With $3$ nodes infected initially, this figure shows the calculation results of (\ref{eq_birth_death_markov_chain_differential_equation}) as Fig. \ref{one_initial_infected} and \ref{two_initial_infected}.]{
      \centering
      \includegraphics[width=152pt]{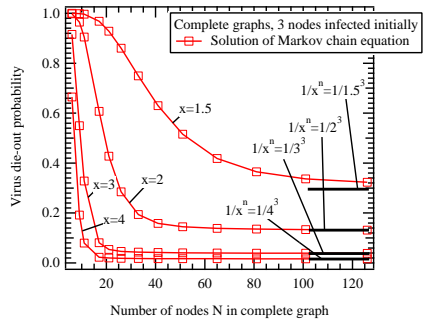}
      \label{three_initial_infected}
    }
    \caption{The virus die-out probability in complete graphs.}\label{fig_die-out_probability}
\end{figure}

After solving the epidemic process (\ref{eq_birth_death_markov_chain_differential_equation}) in the complete graph $K_{126}$ with effective infection rate $\tau=0.016$, Fig. \ref{K126example} shows the prevalence $y(t)$ and the die-out probability $\Pr[S(t)=0]$ as an example. The metastable state is reached approximately at time $t=10$ and hereafter, and the prevalence $y(t)$ keeps steady. Also, the die-out probability $\Pr[S(t)=0]$ becomes approximate constant earlier from $t=5$. The prevalence $y(t)$ decreases slowly to $0$ after an infinitely long time \cite{van_de_bovenkamp_survival_2015,van_mieghem_decay_2013}, and correspondingly, the die-out probability increases to $1$. At $t=45$ in the metastable state, the number of die-out realizations of the SIS epidemic simulation and the solution of the Markov chain Eq. (\ref{eq_birth_death_markov_chain_differential_equation}) are recorded and shown in Fig. \ref{one_initial_infected}, \ref{two_initial_infected}, and \ref{three_initial_infected}. The simulation results in Fig. \ref{one_initial_infected} and \ref{two_initial_infected} are obtained by the SSIS simulator \cite{bovenkamp_epidemic_2015} which applies a Gillespie-like algorithm \cite{gillespie_exact_1977}, and $10^6$ realizations of the Markovian epidemic process are simulated. By counting the number of realizations which have zero infected nodes at $t=45$, the die-out probability is obtained.

Figure \ref{one_initial_infected} and \ref{two_initial_infected} illustrate that, our simulation results match with the computation of the birth-and-death process (\ref{eq_birth_death_markov_chain_differential_equation}). To avoid redundancy, we omit the simulation results in Fig. \ref{three_initial_infected}. From Fig. \ref{one_initial_infected}, the die-out probability at $t=45$ is approximately $1$ corresponding to formula $(\ref{eq_formula_of_die_out_probability})$, when the normalized effective infection rate $x=1$. Also, if $x=1$, the infection rate is below the threshold, and no matter how many nodes are infected initially, the prevalence $y(t)$ decreases exponentially fast for sufficiently large time. The mean-field approximations are usually not accurate around threshold \cite{li_susceptible-infected-susceptible_2012}, which is also verified by Eq. (\ref{eq_S_and_Sq}) when $x=1$ and the die-out probability $\Pr[S(t_m)=0]=1$. For a different number of initially infected nodes $n$, Fig. \ref{fig_die-out_probability} shows that the virus die-out probabilities converge to the concise formula (\ref{eq_formula_of_die_out_probability}) fast with the network size $N$. Furthermore, the larger the normalized effective infection rate $x$ is, the faster the probabilities convergence towards (\ref{eq_formula_of_die_out_probability}).


\subsection{General Graphs}
\label{sec_general}
\begin{figure}[t]
    \subfigure[The virus die-out probability of the SIS epidemic process in an ER graph with the link generation probability $0.9$. The virus spreads starts from $1$, $2$, or $3$ nodes initially.]{
      \centering
      \includegraphics[width=152pt]{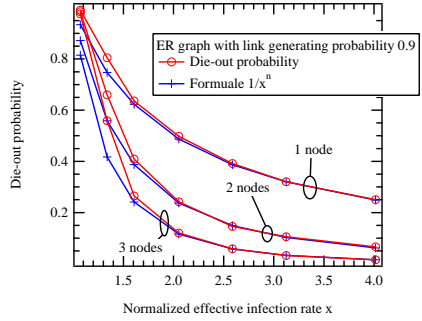}
      \label{ERpi09dieout}
    }
    \qquad
    \centering
    \subfigure[The die-out probability of the SIS epidemic process in another ER graph with the link generation probability $0.5$.]{
      \centering
      \includegraphics[width=152pt]{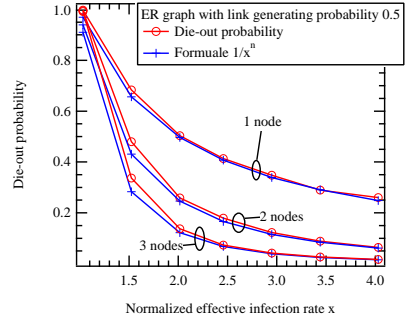}
      \label{ERpi05dieout}
    }
    \\
    \subfigure[The die-out probability of the SIS epidemic process in another ER graph with the link generation probability only $0.1$.]{
      \centering
      \includegraphics[width=150pt]{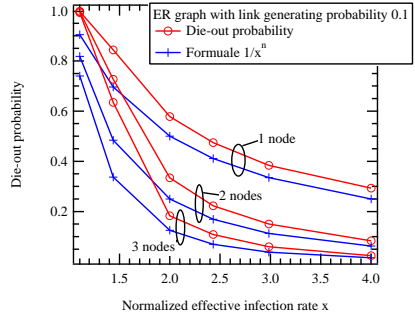}
      \label{ERpi01dieout}
    }
    \qquad
    \centering
    \subfigure[The die-out probability of the SIS epidemic process in a power-law graph.]{
      \centering
      \includegraphics[width=150pt]{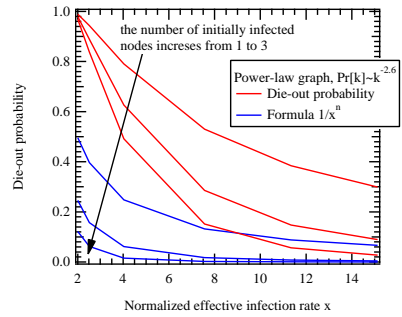}
      \label{BApowerlaw}
    }
    \caption{The virus die-out probability in ER graphs and a power-law graph with different number of initially infected nodes.}
    \label{fig_randomgraphdieout}
\end{figure}

For general graphs, it is infeasible to obtain the virus die-out probability by solving the differential equations of Markov chain, because the number of equations is $2^N$. However, it is still possible to obtain the virus die-out probability efficiently by simulation. We construct three Erd\H{o}s-R\'enyi (ER) graphs $G_p(N)$ with the network size $N=100$ and the link generation probability $p=0.9$, $0.5$, and $0.1$, respectively. The epidemic process is simulated on the ER graphs by randomly choosing the initially infected nodes. For every normalized infection rate $x$ and every number of initially infected nodes $n$, $10^4$ realizations are simulated. Fig. \ref{ERpi09dieout}, \ref{ERpi05dieout}, and \ref{ERpi01dieout} give the the comparison between the die-out probabilities and formula (\ref{eq_formula_of_die_out_probability}) for the number of initially infected nodes $n=1,2,3$. Formula (\ref{eq_formula_of_die_out_probability}) is accurate in the general ER graphs, especially when the normalized effective infection rate $x$ is large. The accuracy of formula (\ref{eq_formula_of_die_out_probability}) decreases with decreasing link generation probability $p$ in ER graphs $G_p(N)$.

The die-out probability of the SIS epidemic process in a power-law graph is presented in Fig. \ref{BApowerlaw} with $10^5$ realizations, and formula (\ref{eq_formula_of_die_out_probability}) shows its limitation. The power law graph has $N=1000$ nodes, and the degree distribution is $\Pr[k]\sim k^{-2.6}$. Fig. \ref{BApowerlaw} exhibits that the die-out probability is almost $1$ when the normalized effective rate is around $2$, which also indicates that the real epidemic threshold in the power-law graph is much larger than the NIMFA threshold $1/\lambda_1$. The inaccuracy of formula (\ref{eq_formula_of_die_out_probability}) is affected by the inaccuracy of the NIMFA threshold as mentioned above.

The simulations seem to indicate that formula (\ref{eq_formula_of_die_out_probability}) is always smaller than the actual die-out probability, which may be attributed to the fact that the NIMFA threshold always lower bounds the actual threshold in any network.

\subsection{NIMFA: Corrected for Die-out}
The mean-field approximation methods are usually not accurate when the initial number of infected nodes is small, because the prevalence obtained by mean-field approximations will generally converge to fixed value due to the existence a steady state, no matter what the initial condition is. When a small number of nodes is initially infected, the die-out probability is relatively large. In this section, we will discuss the accuracy of NIMFA as an example. Previously, the accuracy of NIMFA has been studied from a network topology viewpoint \cite{van_mieghem_accuracy_2015}, but in this section, we focus on the influence of the initial condition.

NIMFA \cite{van_mieghem_virus_2009} reduces the computation complexity of a Markovian epidemic process by assuming independency between the state $X_j(t)$ of node $j$ and the state $X_k(t)$ of node $k$, which closes the governing Eq. (\ref{eq_SIS_governing_equation})
\begin{equation}\label{eq_NIMFA}
\frac{\D v_j(t)}{dt}=-\delta v_j(t)+\beta \sum_{k=1}^Na_{kj}v_k(t)-\beta\sum_{k=1}^N a_{kj}v_j(t)v_k(t)
\end{equation}
where $v_j(t)$ denotes the NIMFA infection probability of node $j$ at time $t$. The NIMFA prevalence is similarly derived as
\begin{equation}\label{eq_NIMFA_prevalence}
y^{(1)}(t)=\frac{1}{N}\sum_{j=1}^Nv_j(t)
\end{equation}
The NIMFA prevalence $y^{(1)}(t)$ decreases exponentially fast to $0$ when the infection rate is below the NIMFA threshold $\tau\leq\tau_c^{(1)}$. If the initial condition $y^{(1)}(0)\neq 0$, the NIMFA prevalence $y^{(1)}(t)$ converges to a non-zero value when $\tau>\tau_c^{(1)}$, which is proved in \cite{khanafer_stability_2014}. Thus, NIMFA is conditioned to the case where the virus in the epidemic process will not die-out, and the absorbing state is removed when $y^{(1)}(0)\neq 0$. Based on (\ref{eq_formula_of_die_out_probability}), we propose an approximate virus surviving probability function at the time $t$ as
\begin{equation}\label{eq_surviving_probability_correcting_function}
f(t)=1-\frac{1}{x^n}+\frac{1}{x^n}\mathrm{e}^{-\lambda_1t}
\end{equation}
Equation (\ref{eq_surviving_probability_correcting_function}) is motivated as follows. At time $t=0$ and $y^{(1)}(t)\neq 0$, the virus surviving probability is $1$ and $f(0)=1$, because a curing event happens with zero probability, when the time interval is $0$. Next, simulations seem to indicate that the virus die-out probability decreases exponentially fast to $1/x^n$ in metastable state with a rate $\lambda_1$.

\begin{figure}
\sidecaption[t]
  \centering
  \subfigure[SIS epidemic process in ER graph with network size $N=50$. The effective infection rate is $
  \tau=0.25$]{
      \centering
      \includegraphics[width=162pt]{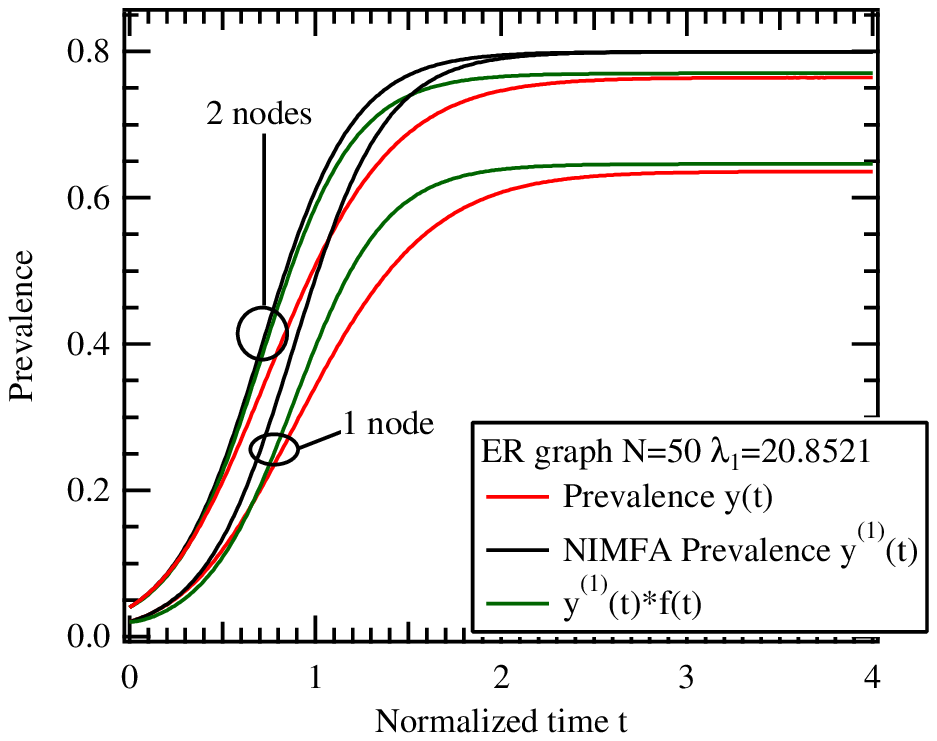}
      \label{ERcorrectNIMFA025}
    }
    \qquad
    \subfigure[SIS epidemic in complete graph $K_{50}$ with effective infection rate $\tau=0.06$]{
      \centering
      \includegraphics[width=164pt]{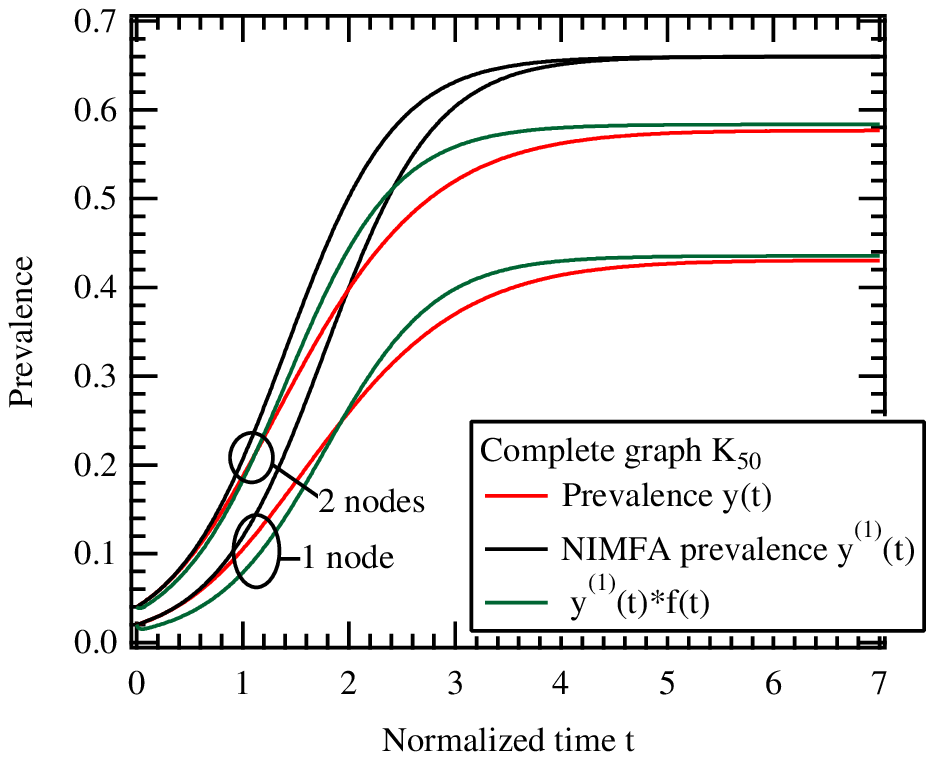}
      \label{K50correctNIMFA006}
    }
  \caption{Comparison of NIMFA and the prevalence. The normalized time is the time scale when the curing rate $\delta=1$ and the prevalence is obtained by averaging $10^6$ realizations.}\label{fig_NIMFA_correct}
\end{figure}

To incorporate the die-out, the NIMFA prevalence can be corrected by applying (\ref{eq_S_and_Sq})
\begin{equation}\label{eq_NIMFA_cerrected_by_ft}
y(t)\approx y^{(1)}(t)f(t)
\end{equation}
Figure \ref{fig_NIMFA_correct} presents the prevalence and the approximation (\ref{eq_NIMFA_cerrected_by_ft}) of the SIS epidemic process in the complete graph $K_{50}$ and the random generated ER graph in Sec \ref{sec_general}. Starting from one or two infected nodes, NIMFA fails to predict the prevalence. The steady state of NIMFA is independent of the initial conditions. Fortunately, (\ref{eq_NIMFA_cerrected_by_ft}) seems a good approximation at the initial stage of the SIS epidemic process.

\section{Conclusion}
In this paper, we discuss the virus die-out probability, which is the probability that the SIS Markovian epidemic process reaches the absorbing state. The importance of the virus die-out probability lies in that it connects the virus spreading phenomena omitting die-out and the exact Markovian model with an absorbing state. Furthermore, we propose an approximate formula (\ref{eq_formula_of_die_out_probability}) of the virus die-out probability, which only contains three essential parameters: the largest eigenvalue of adjacency matrix $\lambda_1$ (the topology parameter), the effective infection rate $\tau$ (the spreading ability parameter), and the number of initially infected node $n$ (the initial condition parameter). If a few nodes are infected, then formula (\ref{eq_formula_of_die_out_probability}) indicates that the virus will almost surely cause a disease outbreak when the infection rate is above the threshold, irrespective of the network size $N$. However, the accuracy of formula (\ref{eq_formula_of_die_out_probability}) also depends on the accuracy of the NIMFA epidemic threshold $1/\lambda_1$. Based on formula (\ref{eq_formula_of_die_out_probability}), an approximate virus surviving probability function (\ref{eq_surviving_probability_correcting_function}) is proposed. We also discuss the correction for NIMFA.

\begin{acknowledgement}
Q. Liu would like to thank the support from China Scholarship Council.
\end{acknowledgement}
\section*{Appendix}
\addcontentsline{toc}{section}{Appendix}
In the gambler's ruin problem, the goal of the virus is to infect a certain number of nodes and to successfully reach the metastable state. If the virus cannot achieve the goal, the virus loses the game and dies out in the network. The analytic solution of the gambler's ruin probability of a birth-and-process, which gives the probability $\mu_n$ that the virus dies out before infecting all $N$ nodes in a finite time starting from an arbitrary number of infected nodes $n$, equals \cite[p.~231]{van_mieghem_performance_2014},
\begin{equation}\label{eq_gambler_ruin}
  \mu_n=\frac{\sum_{k=n}^{N-1}\prod_{m=1}^{k}\frac{1}{(N-m)\tau}}{1+\sum_{k=1}^{N-1}\prod_{m=1}^{k}\frac{1}{(N-m)\tau}}
\end{equation}
First, we evaluate the expression (\ref{eq_gambler_ruin}). Since%
\[
\prod_{m=1}^{k}\frac{1}{\left(  N-m\right)  \tau}=\frac{1}{\tau^{k}}%
\frac{\left(  N-k-1\right)  !}{\left(  N-1\right)  !}%
\]
we have that%
\[
\sum_{k=n}^{N-1}\prod_{m=1}^{k}\frac{1}{\left(  N-m\right)  \tau}=\frac
{1}{\left(  N-1\right)  !}\sum_{k=n}^{N-1}\frac{\left(  N-k-1\right)  !}%
{\tau^{k}}%
\]
Let $j=N-k-1$, then $0\leq j\leq N-n-1$ so that a change of variable results
in%
\[
\sum_{k=n}^{N-1}\frac{\left(  N-k-1\right)  !}{\tau^{k}}=\frac{1}{\tau^{N-1}%
}\sum_{j=0}^{N-n-1}j!\tau^{j}%
\]
Combining all yields (\ref{eq_gambler_ruin_with_polynomial}), it is%
\[
\mu_{n}=\frac{\sum_{j=0}^{N-n-1}j!\tau^{j}}{\sum_{j=0}^{N-1}j!\tau^{j}}%
=\frac{p_{N-n-1}\left(  \tau\right)  }{p_{N-1}\left(  \tau\right)  }%
\]
which is a fraction of two polynomials of the type $p_{m}\left(  z\right)
=\sum_{j=0}^{m}j!z^{j}=1+z+2!z^{2}+\cdots+m!z^{m}$ with positive coefficients
(all derivatives are positive). Thus, $p_{m}\left(  z\right)  $ is rapidly
increasing for $z>0$ and possible real zeros are negative.

The ratio $\frac{j!z^{j}}{\left(  j-1\right)  !z^{j-1}}=jz$ of two consecutive
terms in the polynomial $p_{m}\left(  z\right)  $ indicates that, if $jz>1$
holds for all $1\leq j\leq m$, the terms are increasing, while if $jz<1$ for
all $j$, the terms are decreasing. Hence, if $j\tau<1$ for all $1\leq j\leq
N-1$, which is satisfied if $\tau<\frac{1}{N-1}$, then the terms in
$p_{N-1}\left(  \tau\right)  $ as well as in $p_{N-n-1}\left(  \tau\right)  $
are decreasing and both $p_{N-1}\left(  \tau\right)  $ and $p_{N-n-1}\left(
\tau\right)  $ tend to each other so that $\mu_{n}\rightarrow1$. In the other
case, for $\tau>\frac{1}{N-1}$ and for sufficiently large $N$, the polynomial $p_{N-1}\left(  z\right)$ will be dominated by the largest term and $\mu_{n}$ is approximately equal to%
\begin{align*}
\mu_{n} &  \approx\frac{\left(  N-n-1\right)  !\tau^{N-n-1}}{\left(
N-1\right)  !\tau^{N-1}}=\frac{\left(  N-n-1\right)  !}{\left(  N-1\right)
!}\frac{1}{\tau^{n}}=\frac{1}{\left(  N-1\right)  \left(  N-2\right)
\ldots\left(  N-n\right)  }\frac{1}{\tau^{n}}\\
&  =\frac{1}{\left(  \left(  N-1\right)  \tau\right)  ^{n}\left(  1-\frac
{1}{N-1}\right)  \left(  1-\frac{2}{N-1}\right)  \cdots\left(  1-\frac{n-1}%
{N-1}\right)  }%
\end{align*}
If $n<<N$, then we arrive at formula (\ref{eq_formula_of_die_out_probability})%
\[
\mu_{n}\approx\frac{1}{\left(  \left(  N-1\right)  \tau\right)  ^{n}}%
\]
because $x=\frac{\tau}{\tau_{c}^{\left(  1\right)  }}=\lambda_{1}\tau=\left(
N-1\right)  \tau$ for the complete graph $K_{N}$ as $\lambda_{1}\left(
K_{N}\right)  =N-1$.
\bibliographystyle{spmpsci}
\bibliography{bibitem}
\end{document}